\documentclass[showpacs,preprintnumbers,amsmath,amssymb,epsfig]{revtex4}
\usepackage{graphicx}
\usepackage{dcolumn}
\usepackage{bm}

\begin{document}
\title{Teleportation implementation of non-deterministic quantum logic operations
by using linear optical elements}
\author{XuBo Zou, K. Pahlke and W. Mathis}
\address{Electromagnetic Theory Group at THT,\\
 Department of Electrical
Engineering, University of Hannover, Germany}


\begin{abstract}
We present a feasible scheme to implement the non-deterministic
quantum logic operation of Knill, Laflamme and Milburn (Nature,
409, 46-52(2001)) by a teleportation technique. An entangled
photon channel is generated by emitting two single photons into an
asymmetric beam splitter. In order to generate the classical
channel signal two photon detectors are needed.
\end{abstract}
\pacs{03.67.-a, 03.65.Ud,42.50.-p} \maketitle

The development of quantum algorithms shows impressively, that a
quantum computer can provide an enormous speed up compared to
classical computers \cite{shor} \cite{gro}. This principle
theoretical result motivated an intensive research in the field of
quantum information processing. The idea to develop quantum logic
gates was emerged from reversible Boolean gates \cite{deut}. The
mathematical concept of quantum theory was converted into a
concept of quantum gates which are combined to form quantum
circuits \cite{di}. Apart from system specific investigations on
quantum computer prototypes the interest in principle problems of
using entanglement still remains. Fundamental aspects of quantum
mechanics \cite{daa}, quantum information processing and
communication, cryptography \cite{ds}, quantum teleportation
\cite{db} and quantum dense coding \cite{km} can be very well
investigated with quantum optical systems. The realization of a
quantum computer is particularly appealing because of the robust
nature of quantum states of light against the decoherence.
Recently the implementation of a probabilistic quantum logic gate
was proposed, which is based on linear optical elements
\cite{klm}. This concept uses additional single-photon sources and
single-photon detectors. A non-deterministic gate, which is known
as the nonlinear sign-gate (NLS), was proposed in that paper. This
gate transforms the quantum states, which are labeled by the
photon occupation number, according to
\begin{eqnarray}
\alpha|0\rangle+\beta|1\rangle+\gamma|2\rangle\rightarrow\alpha|0\rangle+\beta|1\rangle-\gamma|2\rangle
\,.\label{1}
\end{eqnarray}
The probability of this outcome is rated to $1/4$. Although the
scheme in Ref \cite{klm} contains only linear optical elements,
the optical network is complex and would present major stability
and mode matching problems in their construction. Several less
complicated schemes with linear optical elements were presented
\cite{ra}\cite{ru} to implement the NLS-gate with a slightly lower
probability of this outcome.\\
If the information is stored in the polarization of photons a
probabilistic quantum logic gate (CNOT-gate and controlled-phase
gate) can also be implemented \cite{bbo}. But this scheme requires
entangled photon states as a resource. Pittman et al. \cite{pi}
proposed an alternative scheme, which uses polarizing beam
splitters (PBS) and the quantum erasure technique. This scheme
makes a high rate of efficiency possible. The authors describe the
implementation of several quantum logic operations of an
elementary nature, including a quantum parity check and a quantum
encoder. They show how these elements can be combined in order to
implement a CNOT-gate. This scheme also requires entangled photon
states as a resource. An experimental demonstration of several
quantum logic devices can
be found in \cite{pi1}.\\
In this paper we present a scheme to implement the
non-deterministic NLS-gate. We adapted this scheme from a
teleportation scheme \cite{db}. To use teleportation schemes for
universal quantum computation was first suggested by Gottesman and
Chuang \cite{chu}.\\
The experimental setup, which we suggest to use, is depicted in
Fig.1. The input quantum state is in the form
\begin{eqnarray}
\Psi_{in}=\alpha|0\rangle_1+\beta|1\rangle_1+\gamma|2\rangle_1\,.
\label{2}
\end{eqnarray}
An ancilla quantum state in the form
$\Psi_{ancilla}=|1\rangle_2|1\rangle_3$ is required. The beam
splitter $BS_1$ transforms this ancilla quantum state into
\begin{eqnarray}
\Psi^{\prime}_{ancilla}=
\frac{\sin2\theta}{\sqrt{2}}(|2\rangle_{2^{\prime}}|0\rangle_{3^{\prime}}-
|0\rangle_{2^{\prime}}|2\rangle_{3^{\prime}})
+\cos2\theta|1\rangle_{2^{\prime}}|1\rangle_{3^{\prime}}\,.
\label{3}
\end{eqnarray}
The reflectance and the transmittance of the beam splitter $BS1$
are denoted by $\sin\theta$ and $\cos\theta$. These quantities
will be determined later. The field state
$\Psi_1\Psi^{\prime}_{ancilla}$ will directly be used for the
NLS-gate implementation. The input modes of Alice's symmetric beam
splitter $BS2$ are the mode $1$ and the mode $2^{\prime}$. Thus,
the quantum state transforms to
\begin{eqnarray}
\Psi_{out}&=&|2\rangle_{1^{\prime\prime}}
|0\rangle_{2^{\prime\prime}}\left[
(\sin2\theta)\alpha|0\rangle_{3^{\prime}}+
2(\cos2\theta)\beta|1\rangle_{3^{\prime}}-
(\sin2\theta)\gamma|2\rangle_{3^{\prime}}\right]/2\sqrt{2}\nonumber\\
&&+|0\rangle_{1^{\prime\prime}} |2\rangle_{2^{\prime\prime}}\left[
(\sin2\theta)\alpha|0\rangle_{3^{\prime}}-
2(\cos2\theta)\beta|1\rangle_{3^{\prime}}-
(\sin2\theta)\gamma|2\rangle_{3^{\prime}}\right]/2\sqrt{2}\nonumber\\
&&+ \Psi_{other}\,. \label{3b}
\end{eqnarray}
The other terms $\Psi_{other}$ of this quantum state don't
contribute to the two events, which we consider in the following.
Alice makes a photon number measurement on the mode
$1^{\prime\prime}$ and the mode $2^{\prime\prime}$ with the
detectors $D_1$ and $D_2$. If $D_1$ detects two photons and $D_2$
does not detect any photon, the state is projected into
\begin{eqnarray}
\Psi_{out}=\frac{1}{2\sqrt{2}}(\sin2\theta\alpha|0\rangle+2\cos2\theta\beta|1
\rangle-\sin2\theta\gamma|2\rangle)\,. \label{4}
\end{eqnarray}
If we choose the reflectance $\sin\theta$ of the beam splitter in
order to fulfill $\sin2\theta=2/\sqrt{5}$, the output quantum
state $\Psi_{out}=\alpha|0\rangle+\beta|1\rangle-\gamma|2\rangle$
is generated from the input quantum state $\Psi_{in}$. This is the
transformation property (\ref{1}) of the NLS-gate. The probability
of this outcome will be $10\%$. If $D_2$ detects two photons and
$D_1$ does not detect any photon, the quantum state is projected
to the form
\begin{eqnarray}
\Psi_{out}=\alpha|0\rangle-\beta|1\rangle-\gamma|2\rangle
\,.\label{5}
\end{eqnarray}
The other terms $\Psi_{other}$ of the quantum state (\ref{3b})
don't contribute to this event. In order to implement the NLS-gate
transformation property (\ref{1}) Bob needs to apply a $\pi$-phase
shifter that changes the signs of the state $|1\rangle$
\cite{woo}. The probability of this outcome will be $10\%$. Hence,
the total success probability of the gate is the probability of
either of these two
outcomes, i.e., $20\%$.\\

A scheme is presented to implement the non-deterministic NLS-gate
with 20 percent of probability of success. Only linear optical
devices in a comparatively simple experimental setup are needed.
Currently available triggered single photon sources operate by
means of fluorescence from a single molecule \cite{cbb} or a
single quantum dot \cite{cs} \cite{pm}. These techniques exhibit a
very good performance. However, in order to generate entangled
photon states a synchronized arrival of many photons at the beam
splitter input ports is needed.\\
The main difficulty of our scheme in respect to an experimental
demonstration consists in the requirement on the sensitivity of
the detectors. These detectors should be capable of distinguishing
between no photon, one photon or two photons. Recently,
experimental techniques for single photon detection made
tremendous progress. A photon detector based on visible light
photon counter can distinguish between a single photon incidence
and two photon incidence. A high quantum efficiency with a good
time resolution and a low bit-error rate was reported \cite{yyy}.
To filfill the requirements of the presented scheme will be
experimentally challenging.\\
Furthermore, this scheme can be employed by means of a symmetric
beam splitter $BS_1$ to implement quantum teleportation of
superpositions of vacuum and two-photon states.

\begin{flushleft}

{\Large \bf Figure Captions}

\vspace{\baselineskip}

{\bf Figure .} The schematic is shown to implement the
non-deterministic quantum logic operation using teleportation
protocol. The entangled channel is generated via a pair of photon
incident on Beam splitter $BS_1$. The wave $3^{\prime}$ is sent to
the Bob, while wave $2^{\prime}$ is sent to Alice who combines
with the wave 1 to be implement. Alice makes a measurement upon
the wave $2^{\prime}$ and wave 1 and information the result to Bob
via a classical communication channel(represent by a dot line).
Based on the result of Alice measurement, Bob can perform a
suitable unitary transformation with a $\pi$ phase shift.

\end{flushleft}

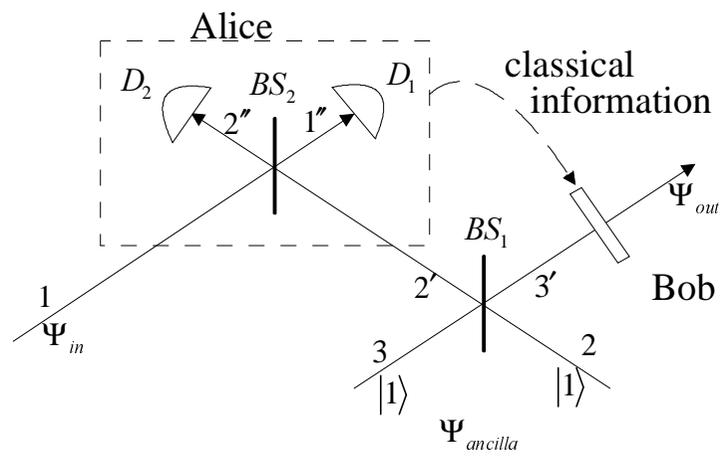
\end{document}